\documentclass[12pt]{article}
\usepackage{graphicx}
\usepackage{amsmath}
\usepackage{amssymb}
\usepackage{caption2}
\begin{document}
\bibliographystyle{prsty}
\begin{center}
{\large {\bf \sc{  Analysis of the vertex   $D^*D^* \rho$ with   the light-cone QCD sum rules  }}} \\[2mm]
Zhi-Gang Wang$^{1}$ \footnote{E-mail,wangzgyiti@yahoo.com.cn.  }, Zhi-Bin Wang$^{2}$     \\
$^{1}$ Department of Physics, North China Electric Power University, Baoding 071003, P. R. China \\
$^{2}$ College of Electrical Engineering, Yanshan University, Qinhuangdao 066004, P. R. China \\
\end{center}

\begin{abstract}
In this article, we analyze the vertex $D^*D^*\rho$ with  the
  light-cone QCD sum rules. The strong coupling constant $g_{D^*D^*\rho}$
  is an important  parameter in evaluating the charmonium absorption
cross sections in searching for the quark-gluon plasmas. Our
numerical value for the  $g_{D^*D^*\rho}$ is consistent with the
prediction  of the effective $SU(4)$ symmetry and vector meson
dominance theory.
\end{abstract}


\section{Introduction}

The suppression of  $J/\psi$ production in  relativistic  heavy ion
collisions maybe  one of the important signatures to identify the
possible phase  transition  to the quark-gluon plasma
\cite{Matsui86}. The dissociation of the $J/\psi$ in the quark-gluon
plasma due to color screening can lead to a reduction of its
production, on the other hand, the $J/\psi$ suppression maybe
already present in the hadron-nucleus collisions. It is necessary to
know absorption of the $J/\psi$  by the co-mover light mesons before
we can make a definitive conclusion.
 The values of the
$J/\psi$ absorption cross sections by the light hadrons  are not
known empirically, we have to resort to some theoretical approaches.
Among existing  approaches,  the one-meson exchange model and the
effective $SU(4)$ theory are typical \cite{MesonEx, SU4}. The
detailed knowledge  about the strong coupling constants which are
basic parameters in the effective Lagrangian is of great importance.
Furthermore, the strong coupling constants among the charmed mesons
and light-mesons  play an important role in understanding
final-state interactions in the hadronic $B$ decays \cite{CHHY}.

 There have been  many works
dealing with the strong coupling constants concerning the charmed
mesons, see e.g. Refs.\cite{gg,Wanggg}.  In Ref.\cite{Nielsen0710},
the authors calculate the strong coupling constant $g_{D^* D^*
\rho}$ with the QCD sum rules, and obtain much larger value than
what expected from the effective $SU(4)$ theory. In this article, we
calculate the value of the $g_{D^* D^* \rho}$ with   the light-cone
QCD sum rules. The light-cone QCD sum rules  carry out the operator
product expansion near the light-cone $x^2\approx 0$ instead of the
short distance $x\approx 0$ while the non-perturbative matrix
elements are parameterized by the light-cone distribution amplitudes
 (which classified according to their twists)  instead of
 the vacuum condensates \cite{LCSR}.

The article is arranged as: in Section 2, we derive the strong
coupling constant  $g_{D^*D^*\rho}$  with  the light-cone QCD sum
rules; in Section 3, the numerical result and discussion; and in
Section 4, conclusion.

\section{Strong coupling constant  $g_{D^*D^*\rho}$  with light-cone QCD sum rules}

We study the strong coupling constant  $g_{D^*D^* \rho}$
 with the two-point correlation function $\Pi_{\mu\nu}(p,q)$,
\begin{eqnarray}
\Pi_{\mu\nu}(p,q)&=&i \int d^4x \, e^{-i q \cdot x} \,
\langle 0 |T\left\{J_\mu(0) J_\nu^+(x)\right\}|\rho(p)\rangle \, , \\
J_\mu(x)&=&{\bar u}(x)\gamma_\mu   c(x)\, ,  \nonumber \\
J_\nu(x)&=&{\bar d}(x)\gamma_\nu   c(x)\, ,
\end{eqnarray}
where  the currents $J_\mu(x)$ and $J_\nu(x)$ interpolate the vector
mesons $D^{*0}$ and $D^{*+}$ respectively, the external state $\rho$
has the four momentum $p_\mu$ with $p^2=m_\rho^2$.

According to the basic assumption of current-hadron duality in the
QCD sum rules \cite{SVZ79}, we can insert  a complete series of
intermediate states with the same quantum numbers as the current
operators  into the correlation function  to obtain the hadronic
representation. After isolating the ground state contributions from
the pole terms of the mesons $D^*$, we get the following result,
\begin{eqnarray}
\Pi_{\mu\nu }(p,q) &=&-\frac{f_{D^*}^2M_{D^*}^2 g_{D^*D^* \rho}}
{\left\{M_{D^*}^2-(q+p)^2\right\}\left\{M_{D^*}^2-q^2\right\}}2g_{\mu\nu}\epsilon
\cdot q    + \cdots    \, ,
\end{eqnarray}
where the following weak decay constant and phenomenological
Lagrangian have been used,
\begin{eqnarray}
\langle0 | J_\mu(0)|D^*(p)\rangle&=&f_{D^*}M_{D^*}\epsilon_\mu\, ,
\nonumber \\
 \mathcal{L}&=& \frac{ig_{ D^{*}D^{*}\rho}}{\sqrt{2}} \left \{
 \partial_{\mu}D^{*}_{\nu} \rho^{\mu} \bar{D}^{*\nu}
-D^{*}_{\nu}\rho_{\mu}\partial^{\mu}\bar{D}^{*\nu}+
D^{*}_{\nu}\partial_{\mu} \rho^{\nu}\bar{D}^{*\mu}
 \right.\nonumber\\
&&\left. - \partial_{\mu}
D^{*}_{\nu}\rho^{\nu}\bar{D}^{*\mu}+D^{*}_{\mu}
\rho_{\nu}\partial^{\mu}\bar{D}^{*\nu} -D^{*}_{\mu}
\partial^{\mu}\rho_{\nu} \bar{D}^{*\nu}  \right\} \, ,
\end{eqnarray}
here $\rho=\sigma^i \rho_i$, $D^*=(D^{*0},D^{*+})$ \cite{MesonEx,
SU4}.

We carry out the  operator product expansion for the correlation
function $\Pi_{\mu\nu }(p,q)$ in perturbative QCD theory, and obtain
 the analytical expressions at the level of quark-gluon degrees of freedom,
 then perform the double Borel transformation
and match the  quark-hadron duality below the threshold $s_0$,
finally we obtain the sum rule for the strong coupling
  constant $g_{D^*D^*\rho}$,
\begin{eqnarray}
&&2g_{D^*D^* \rho}f_{D^*}^2 M^2_{D^*}   \exp\left\{-\frac{M_{D^*}^2}{M^2_1}-\frac{M^2_{D^*}}{M_2^2}\right\}\nonumber\\
&=& f_{\rho} m_{\rho} M^2\phi_\parallel(u_0)\left\{\exp\left[-
\frac{m_c^2+u_0(1-u_0)m_{\rho}^2}{M^2}
\right]-\exp\left[- \frac{s^0_{\rho}}{M^2} \right] \right\} \nonumber\\
&&+\exp\left[- \frac{m_c^2+u_0(1-u_0)m_{\rho}^2}{M^2} \right]\left\{
\left[f_{\rho}^\perp-f_{\rho}\frac{m_u+m_d}{m_{\rho}}\right]m_c
m_{\rho}^2
 h_{||}^{(s)}(u_0) \right.\nonumber\\
&&\left.-\frac{f_{\rho} m_{\rho}^3A(u_0)}{4}  \left[1
+\frac{m_c^2}{M^2}\right] -  \frac{2f_{\rho}m_c^2 m_{\rho}^3}{M^2}
\int_0^{u_0} d\tau \int_0^\tau dt C(t)\right\} \, ,
\end{eqnarray}
where
\begin{eqnarray}
u_0&=&\frac{M_1^2}{M_1^2+M_2^2}\, , \nonumber \\
M^2&=&\frac{M_1^2M_2^2}{M_1^2+M_2^2} \, .
\end{eqnarray}
Here we have neglected the contributions from the gluons, the
contributions proportional to $G_{\mu\nu}$ can give rise to
three-particle (and four-particle) meson distribution amplitudes
with a gluon (and quark-antiquark pair), their corrections are
always suppressed by the large Borel parameter and  will not exceed
$20\%$, for examples, one can consult the last two articles of
Ref.\cite{Wanggg}. In calculation, the two-particle $\rho$ meson
light-cone distribution amplitudes up to twist-4 have been used, for
explicit expressions, one can consult Ref.\cite{VMLC}. Due to the
special tensor structure $g_{\mu\nu}\epsilon \cdot q$, some
two-particle twist-4 light-cone distribution amplitudes have no
contributions.  The parameters in the light-cone distribution
amplitudes are scale dependent, in this article,  we take
$\mu=1\rm{GeV}$. As the dominating contribution (about $90\%$) comes
from the two-particle twist-2 term involving the
$\phi_\parallel(u)$, for other terms, the continuum subtractions
will not affect the result remarkably, we can neglect the
subtractions.  In some cases, the contributions from the
two-particle twist-3
 light-cone distribution amplitudes are very large (or dominating)
\cite{Wangform}, they depend heavily on the currents we choose to
interpolate the mesons.

\section{Numerical result and discussion}
The input parameters are taken as $m_c=(1.35\pm 0.10)\rm{GeV}$,
$m_u=m_d=(0.0056\pm0.0016)\rm{GeV}$,
$f_\rho=(0.216\pm0.003)\rm{GeV}$,
$f_\rho^{\perp}=(0.165\pm0.009)\rm{GeV}$,
 $m_\rho=0.775\rm{GeV}$,
 $a_1^{\parallel}=0.0$, $a_1^{\perp}=0.0$,
$a_2^{\parallel}=0.15\pm0.07$,  $a_2^{\perp}=0.14\pm0.06$,
$\zeta^{\parallel}_3=0.030\pm 0.010$,
$\widetilde{\lambda}_3^{\parallel}=0.0$,
$\widetilde{\omega}_3^{\parallel}=-0.09\pm 0.03$,
$\kappa_3^{\parallel}=0.0$,  $\omega_3^\parallel=0.15\pm0.05$,
$\lambda_3^\parallel=0.0$,  $\kappa_3^\perp=0.0$,
$\omega_3^\perp=0.55\pm0.25$,  $\lambda_3^\perp=0.0$,
$\zeta_4=0.15\pm 0.10$, $\zeta_4^T=0.10\pm 0.05$ and
$\widetilde{\zeta}_4^T=-0.10\pm 0.05$ \cite{VMLC}. The
 central value of the decay
constant $f_{D^*}$ from lattice simulation is about
$f_{D^*}=0.23\rm{GeV}$ \cite{decayCV}. In this article, we take the
value $f_{D^*}=(0.22 \pm 0.02)\rm{GeV}$ from the two-point QCD sum
rules without perturbative $\mathcal {O}(\alpha_s)$ corrections  for
consistency \footnote{One can consult the last article  of
Ref.\cite{SVZ79} or Ref.\cite{Threshold} for the explicit
expression.}.
 The duality threshold parameter  $s_0$ is chosen to be $s^0_\rho=(6.5\pm0.5)\rm{GeV}^2$,
 the numerical (central) value of $s_0$ is
taken from the QCD sum rules for the mass of the $D^{*}$
\cite{Threshold}. The Borel parameters are chosen as $ M_1^2=M_2^2$
and $M^2=(3-7)\rm{GeV}^2$, in those regions, the value of the
$g_{D^*D^*\rho}$ is   rather stable, the uncertainty from the Borel
parameter is very small, less than $5\%$.

In the limit of large Borel parameter $M^2$, the strong coupling
constant $g_{D^*D^*\rho}$   takes up the following behavior,
\begin{eqnarray}
g_{D^*D^*\rho} &\propto&  \frac{M^2f_{\rho}
\phi_\parallel(u_0)}{f_{D^*}^2}\propto  \frac{M^2f_{\rho}
a^\parallel_2}{f_{D^*}^2} \, .
\end{eqnarray}
It is not unexpected, the contribution  from the twist-2 light-cone
distribution amplitude  $\phi_\parallel(u)$ is greatly enhanced by
the large Borel parameter $M^2$, its contribution is dominating,
about $90\%$, (large) uncertainties of the relevant parameters
presented in above equation have significant impact on the numerical
result. The main uncertainties come from the two parameters
 $f_{D^*}$ and $a^\parallel_2$,   variations of
the two parameters can lead to large uncertainties, about
$(10-20)\%$.

Taking into account all the uncertainties, finally we obtain the
numerical value for  the strong coupling constant $g_{D^*D^*\rho}$ ,
which is shown in Fig.1,
\begin{eqnarray}
  g_{D^*D^*\rho} &=&2.6 \pm 0.7 \, .
\end{eqnarray}
If we take the replacement $\exp\left[-
\frac{m_c^2+u_0(1-u_0)m_{\rho}^2}{M^2} \right] \rightarrow
\exp\left[- \frac{m_c^2+u_0(1-u_0)m_{\rho}^2}{M^2}
\right]-\exp\left[- \frac{s^0_{\rho}}{M^2} \right] $ to subtract the
continuum contributions of the terms besides  the twist-2 light-cone
distribution amplitude, the central value $g_{D^*D^*\rho} =2.6$ will
decrease  about $5\%$.
 Comparing with the values from the effective $SU(4)$ symmetry
and vector meson dominance theory, $g_{D^*D^*\rho}=g_{DD\rho}=2.52$
\cite{SU4}, our numerical value $g_{D^*D^*\rho}\rightarrow
\frac{g_{D^*D^*\rho}}{\sqrt{2}}=1.8 \pm 0.5 $ is much smaller. In
the vector meson dominance theory, $g_{DD\rho}=
\frac{m_\rho}{\sqrt{2}f_\rho}$, the contributions from the radial
excited states $\rho(1450)$, $\rho(1700)$, $\rho(1900)$, $\cdots$
are neglected. We can make a crude estimation with the simple
replacement $m_\rho\rightarrow \frac{m_\rho m_{1450}}{m_\rho+
m_{1450}}$ to take into account the contribution from the
$\rho(1450)$, $g_{D^*D^*\rho}=g_{DD\rho}=1.66$, our numerical result
is reasonable, the $SU(4)$ symmetry breaking effect for the strong
coupling constants is small. In Ref.\cite{Nielsen0710}, the authors
introduce some functions to parameterize the form-factor for small
spacelike  $Q^2$, then extrapolate to the mass shell, much larger
value is obtained, $g_{D^*D^*\rho} =6.6 \pm 0.3 $. Although the
model functions have solid theoretical foundation at large $Q^2$,
extrapolation to the mass shell may have good or bad behaviors,
which correspond to the systematic errors. It is not unexpected
that the values from the QCD sum rules and light-cone QCD sum rules
are different.

\begin{figure}
 \centering
 \includegraphics[totalheight=8cm]{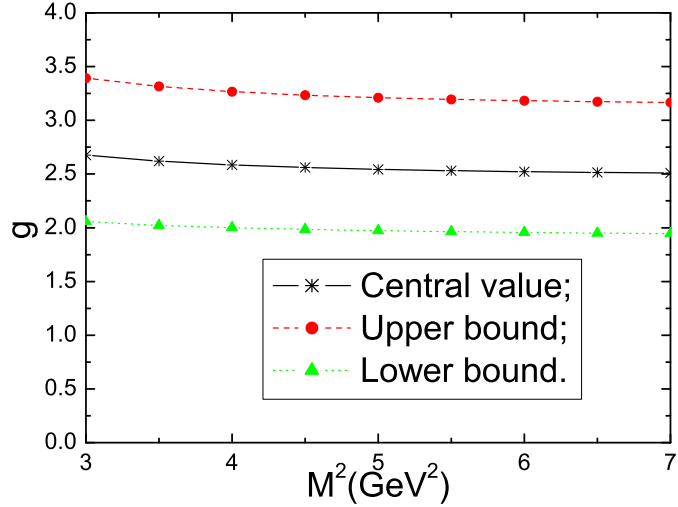}
 \caption{The   $g_{D^*D^*\rho}$ with the Borel parameter $M^2$. }
\end{figure}

\section{Conclusion}

In this article, we analyze the vertex $D^*D^*\rho$ with  the
  light-cone QCD sum rules. The strong coupling constant $g_{D^*D^*\rho}$
  is an important  parameter in evaluating the charmonium absorption
cross sections in searching for the quark-gluon plasmas. Our
numerical value for the  $g_{D^*D^*\rho}$ is consistent with the
prediction of the effective $SU(4)$ symmetry and vector meson
dominance theory.

\section*{Acknowledgments}
This  work is supported by National Natural Science Foundation,
Grant Number 10405009, 10775051, and Program for New Century
Excellent Talents in University, Grant Number NCET-07-0282, and Key
Program Foundation of NCEPU.


\begin{thebibliography}{99}

\bibitem{Matsui86} T. Matsui, H. Satz,  Phys. Lett. {\bf B178} (1986)
416; R. Vogt, Phys. Rept. {\bf 310} (1999) 197.

\bibitem{MesonEx} S. G. Matinyan, B. Muller, Phys. Rev. {\bf C58} (1998)
2994; K. L. Haglin, Phys. Rev. {\bf C61} (2000) 031902; Z. W. Lin,
C. M. Ko, Phys. Rev. {\bf C62} (2000) 034903; A. Sibirtsev, K.
Tsushima, A. W. Thomas, Phys. Rev. {\bf C63} (2001) 044906.



\bibitem{SU4} Y. s. Oh, T. Song, S. H. Lee, Phys. Rev. {\bf C63} (2001) 034901.

\bibitem{CHHY}   H. Y. Cheng, C. K. Chua, A. Soni,
Phys. Rev. {\bf D71} (2005) 014030.

\bibitem{gg}  F. S. Navarra, M. Nielsen, M. E. Bracco, Phys .Rev. {\bf D65} (2002)
037502; Z. H. Li, T. Huang, J. Z. Sun, Z. H. Dai, Phys. Rev. {\bf
D65} (2002) 076005; M. E. Bracco, M. Chiapparini, F. S. Navarra, M.
Nielsen, Phys. Lett. {\bf B605} (2005) 326.

\bibitem{Wanggg} Z. G. Wang, S. L. Wan, Phys. Rev. {\bf D74} (2006)
014017; Z. G. Wang, Eur. Phys. J. {\bf C52} (2007) 553; Z. G. Wang,
Nucl. Phys. {\bf A796} (2007) 61.

\bibitem{Nielsen0710} M. E. Bracco, M. Chiapparini, F. S.
Navarra, M. Nielsen, arXiv:0710.1878[hep-ph].


\bibitem{LCSR}
I. I. Balitsky, V. M. Braun and A. V. Kolesnichenko, Nucl. Phys.
{\bf B312} (1989) 509;  V. L. Chernyak and I. R. Zhitnitsky, Nucl.
Phys. {\bf B345} (1990) 137;  V. M. Braun and I. E. Filyanov, Z.
Phys.  {\bf C44} (1989) 157; V. M. Braun and I. E. Filyanov, Z.
Phys. {\bf C48} (1990) 239.

\bibitem{SVZ79} M. A. Shifman, A. I. and Vainshtein and V. I. Zakharov,
Nucl. Phys. {\bf B147} (1979) 385, 448; L. J. Reinders, H.
Rubinstein and S. Yazaky, Phys. Rept. {\bf 127} (1985) 1.

\bibitem{VMLC} P. Ball, V. M. Braun, Nucl. Phys. {\bf B543} (1999) 201;
 P. Ball, V. M. Braun, Phys. Rev. {\bf D54} (1996)
2182; P. Ball, V. M. Braun, Y. Koike, K. Tanaka, Nucl. Phys. {\bf
B529} (1998) 323; P. Ball, G. W. Jones, R. Zwicky, Phys. Rev. {\bf
D75} (2007) 054004; P. Ball, G. W. Jones, JHEP {\bf 0703} (2007)
069.

\bibitem{Wangform} Z. G. Wang, J. Phys. {\bf G34} (2007) 2183; Z. G. Wang and S. L. Wan,
Eur. Phys. J. {\bf C50} (2007) 781; Z. G. Wang and Z. B. Wang,
arXiv:0711.2921[hep-ph].


\bibitem{decayCV} K. C. Bowler et al, Nucl. Phys. {\bf B619} (2001) 507.

\bibitem{Threshold} A. Hayashigaki, K. Terasaki, hep-ph/0411285.



\end{thebibliography}
\end{document}